# EQUIP-RPT: A framework for objective evaluation of quantitative imaging methods in radiopharmaceutical therapies

Zekun Li[1]; Carlos F. Uribe[2]; Fereshteh Yousefirizi[2]; Lidia Strigari[3]; Neeta Pandit-Taskar[4]; Jazmin Schwartz[5]; Julia Brosch-Lenz[6]; Babak Saboury[7]; Joyita Dutta[8]; Tyler J. Bradshaw[9]; Habib Zaidi[10]; Ronald Boellaard[11]; Kuangyu Shi[12]; M. Soltani[13]; Arman Rahmim[14]; Abhinav K. Jha[15*]

1. School of Biomedical Engineering, ShanghaiTech University, Shanghai, China
2. Department of Basic & Translational Research, BC Cancer Research Institute, Vancouver, BC, Canada
3. Department of Medical Physics, IRCCS Azienda Ospedaliero-Universitaria di Bologna, Bologna, Italy
4. Department of Radiology, University of Pittsburgh School of Medicine, Pittsburgh, PA, USA
5. Long Island Jewish Medical Center, Northwell Health, Department of Radiology, Division of Nuclear Medicine, New Hyde Park, New York, United States
6. Nuclear Medicine, Faculty of Medicine, University of Augsburg, Augsburg, Germany
7. Department of Computational Nuclear Oncology, Institute of Nuclear Medicine, Bethesda, MD, USA
8. Department of Biomedical Engineering, University of Massachusetts Amherst, Amherst, MA, USA
9. Department of Radiology, University of Wisconsin–Madison, Madison, WI, USA
10. Division of Nuclear Medicine and Molecular Imaging, Geneva University Hospitals (HUG), Geneva, Switzerland
11. Department of Radiology and Nuclear Medicine, Amsterdam UMC, Vrije Universiteit Amsterdam, Amsterdam, The Netherlands
12. Department of Nuclear Medicine, Inselspital, Bern University Hospital, University of Bern, Bern, Switzerland
13. Department of Mechanical Engineering, K. N. Toosi University of Technology, Tehran, Iran
14. Departments of Radiology and Physics, University of British Columbia, Vancouver, BC, Canada
15. Department of Biomedical Engineering and Mallinckrodt Institute of Radiology, Washington University, St. Louis, MO, USA

Email: a.jha@wustl.edu

**Abstract**

**Purpose**

Quantitative imaging (QI) provides a non-invasive approach for dosimetry and quantitative response assessment in radiopharmaceutical therapies (RPTs). Multiple QI methods, including emerging physics and artificial intelligence-based methods in instrumentation, acquisition, reconstruction, and image analysis, are being developed. However, a key barrier to routine adoption of these methods is that evidence demonstrating reliable performance of QI methods on clinically relevant quantification tasks in RPTs remains limited. Objective task-based evaluation is therefore needed to evaluate QI methods for RPTs and to build such an evidence base. Our goal is to outline the Evaluation of Quantitative Imaging Performance in Radiopharmaceutical Therapy (EQUIP-RPT) framework, a structured approach for objective, task-based evaluation of QI methods for RPTs.

**Approach**

We outline the EQUIP-RPT framework based on image-science and task-based image-quality evaluation methodology, practical considerations, and recent literature. The framework is organized around four linked design steps: (i) specification of the quantification task, (ii) design of the study population and imaging protocol, (iii) task-specific information extraction, and (iv) selection of figures of merit (FoMs).

**Results**

The EQUIP-RPT framework is presented in the context of four quantification tasks in RPT: activity uptake, absorbed dose, volumetric tumor burden, and lesion-level longitudinal quantifications. The framework distinguishes two paradigms for task-specific information extraction and organizes FoMs according to the performance attributes they characterize. We describe key elements to apply the EQUIP-RPT framework in simulation, physical phantom, and clinical studies. Finally, we show how EQUIP-RPT can be used to generate a quantitative claim based on the evaluation study.

**Conclusion**

The EQUIP-RPT framework is designed to provide a structured approach for objective task-based evaluation of QI methods in RPTs, thereby facilitating clinical translation of these methods.

## 1. Introduction

Radiopharmaceutical therapies (RPTs), including those based on lutetium-177, actinium-225, and lead-212, have attracted growing interest as effective treatments of disseminated malignancies[1]. In RPTs, radiolabeled compounds selectively localize and deliver cytotoxic radiation to lesions, typically through alpha- or beta-particle emissions, inducing DNA damage and leading to localized cell death. However, these isotopes may also accumulate in radiosensitive organs, causing off-target toxicity[1]. Accordingly, an important need exists for methods to accurately and precisely estimate absorbed dose delivered to both lesions and organs at risk. Such quantitative dosimetry is essential for adapting treatment regimens, monitoring and mitigating adverse events, and predicting therapy outcomes, thereby supporting personalized RPTs[2].

Quantitative imaging (QI) provides a non-invasive approach for estimating activity uptake distributions needed for dosimetry and quantitative response assessment in RPTs. Single-photon emission computed tomography (SPECT) detects gamma-ray photons emitted by radiopharmaceuticals and can enable imaging-based activity uptake quantification, dosimetry, and response assessment[3]. Positron emission tomography (PET) with matched or target-specific tracers can provide quantitative biomarkers that inform clinical decision-making. Further, pre-therapy PET may also provide surrogate estimates of radiopharmaceutical uptake, thereby assisting activity prescription and dose planning[4].

Given this need, multiple new QI methods, including those for instrumentation, acquisition protocols, reconstruction, and analysis, are being developed. For instrumentation, cadmium–zinc–telluride (CZT) SPECT systems and systems with stationary or other novel geometries may offer improved energy resolution and sensitivity compared with conventional systems[5]. Long-axial-field-of-view, or total-body, PET provides substantially higher sensitivity and coverage than conventional systems and may improve the precision of PET-derived quantitative biomarkers for RPTs[6]. Advances in improving acquisition protocols include optimizing energy-window definitions, collimator selection, and sampling schedules for QI in RPTs[7,8]. In reconstruction and quantification, developments include physics-based system modelling and projection-domain methods that directly estimate activity uptake from projection data[9]. Artificial intelligence (AI)-based methods, including those for image enhancement, automated segmentation, data-driven partial-volume correction, and surrogate dose prediction, have also reported improvements in quantification performance[10].

Alongside these developments, QI is recommended and applied clinically in several RPTs. One example is SPECT-based dosimetry in multiple lutetium-177-based therapies[3,11]. However, routine QI implementation and dosimetry-guided treatment remain limited in RPTs[12]. One barrier to broader adoption is insufficient evidence demonstrating reliable performance of QI methods on clinically relevant quantification tasks in RPTs. Establishing such evidence is challenging because these tasks typically involve multistep pipelines, where each stage has its own challenges and errors introduced at one stage can propagate downstream.

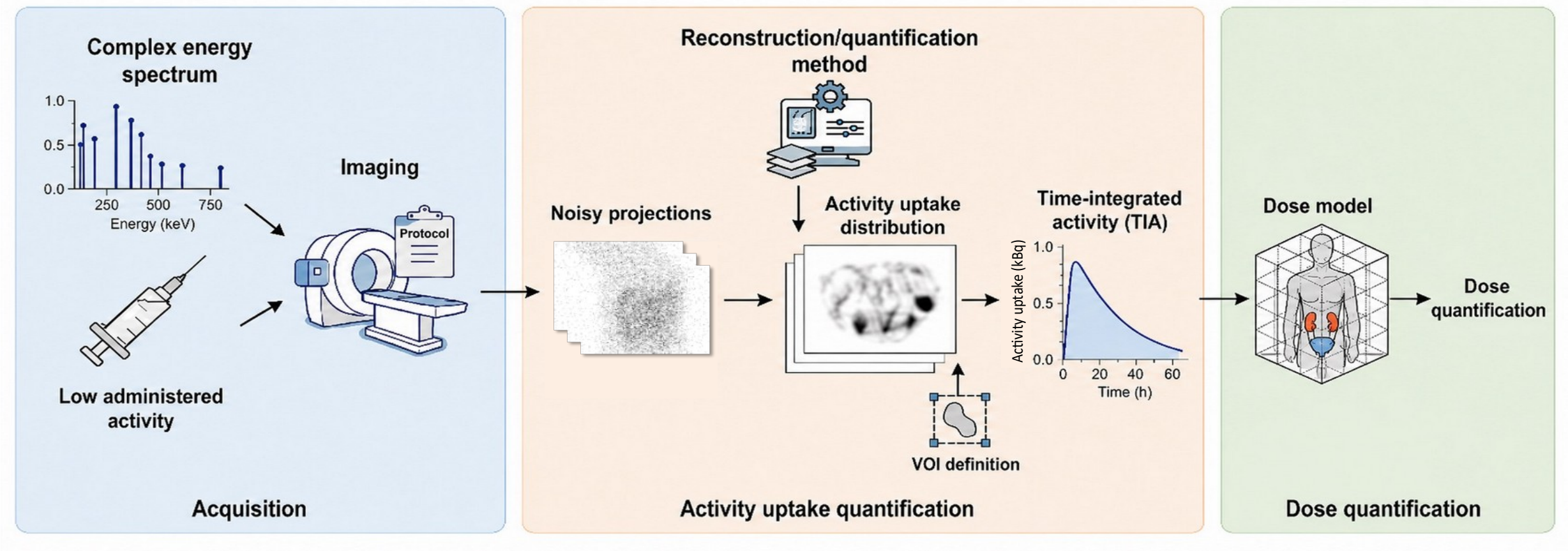

Figure 1. Illustration of the workflow for SPECT-based dose quantification in RPTs.

To illustrate this, consider a representative pipeline for SPECT-based absorbed dose quantification (Fig. 1). At the acquisition stage, complex decay chains, multi-peak emission spectra with high-energy emissions, and image-degrading processes such as attenuation, scatter and collimator-detector effects can degrade the quality of the acquired data[8]. For alpha-particle-emitting RPTs, low administered activities can lead to low detected counts and noisy measurements[8]. If these effects are insufficiently compensated, they propagate to reconstruction and bias downstream activity and dose estimates.

In reconstruction, conventional parameter settings are often selected for generic noise-resolution trade-offs or visual appearance rather than clinically relevant quantitative tasks[13,14]. For example, post-filtering, which is commonly applied in SPECT reconstruction, can improve visual smoothness but reduce the accuracy of activity quantification and thus bias subsequent dose quantification[14]. Regional uptake quantification from reconstructed images then requires VOI definition and image segmentation, both of which can introduce systematic errors and operator-dependent variability[15]. Partial-volume effects, due to the limited spatial resolution of PET and SPECT, further bias activity estimates, particularly for small regions[16].

For absorbed-dose estimation, additional bias and variance are introduced by the temporal dimension of the measurements. Multi-time-point dosimetry is sensitive to image registration across time points, variability in sampling schedules, and pharmacokinetic modelling choices. Reduced-time-point strategies can reduce patient burden and simplify workflow, but may affect both accuracy and precision and increase sensitivity to deviations from the schedule and modelling assumptions[17]. Differences in dose-calculation models and reporting levels can further contribute to variability in absorbed-dose estimates[15,18].

Reproducibility across sites, scanners, and analysis pipelines is another major issue. For example, in the Society of Nuclear Medicine and Molecular Imaging (SNMMI) lutetium-177 dosimetry challenge, between-center dispersion in absorbed-dose estimates reached a quartile coefficient of dispersion up to 73% for liver in a representative case, and this variability was substantially reduced when part of upstream variability in the quantification pipeline was removed[15,18]. These findings show that differences in processing setup within the pipeline can significantly vary the quantification results. Thus, evaluations demonstrating reliable performance of QI methods at individual stages of the pipeline are important, but by themselves are insufficient to establish reliable performance on the final clinical quantification task.

These considerations indicate that clinical implementation of QI for RPT will require not only continued methodological innovation but also techniques that can reliably quantify performance on clinically relevant quantification tasks. A rigorous way to address this need is to frame the evaluation of QI methods within the paradigm of objective task-based image-quality evaluation[19,20]. In this paradigm, imaging systems and algorithms are quantitatively assessed according to their performance on the clinical tasks under clinically relevant experimental settings. This paradigm is well established for detection and discrimination tasks, where it has provided a quantitative framework for designing and optimizing imaging methods that achieve reliable task performance[13,21]. However, their application to quantification tasks has only recently received broader attention[9,22,23]. Within this paradigm, QI methods are assessed using prespecified, clinically meaningful quantification tasks in a rigorously designed study[13,19,22,24,25]. As a concrete example, when evaluating a new SPECT system for organ-level activity quantification, the relevant performance measures are not generic image-quality scores but rather figures of merit (FoMs) that quantify the reliability of organ activity estimates from a prespecified estimation pipeline across a representative and clinically realistic patient population.

Conducting such evaluations in RPT requires a structured study-design framework tailored to prespecified quantification tasks, multistep QI pipelines in RPT, and range of evaluation settings available to investigators. In this work, we propose the Evaluation of Quantitative Imaging Performance in Radiopharmaceutical Therapy (EQUIP-RPT) framework and associated practical considerations for objective evaluation of QI methods in RPT. The EQUIP-RPT framework is informed by image-science methodology, practical experience, and recent literature. Existing guidance from the Medical Internal Radiation Dose (MIRD) Committee and the European Association of Nuclear Medicine (EANM) provides technical recommendations for performing QI in RPTs[7,11,14]. EQUIP-RPT complements these guidelines by providing a framework for objective task-based evaluations of QI methods for RPTs. Section 2 introduces the four steps of EQUIP-RPT and shows how these steps are combined to define an explicit evaluation claim: a statement of QI method performance for a quantification task in RPT under specified study setups. Section 3 describes how evaluations can be implemented in simulation, physical phantom, and clinical studies. Rather than prescribing a single protocol, EQUIP-RPT provides a structured set of design choices that supports objective evaluation of emerging QI methods in RPT.

## 2. The EQUIP-RPT Framework for Objective Task-Based Evaluation

We introduce the EQUIP-RPT framework for objective task-based evaluation of QI methods for quantification tasks in RPTs. As illustrated in Figure 2, EQUIP-RPT comprises four design steps:

1. Specification of the Quantification Task
2. Design of Study Populations and Imaging Protocols
3. Task-Specific Information Extraction
4. Selection of figures of merit to quantify task performance

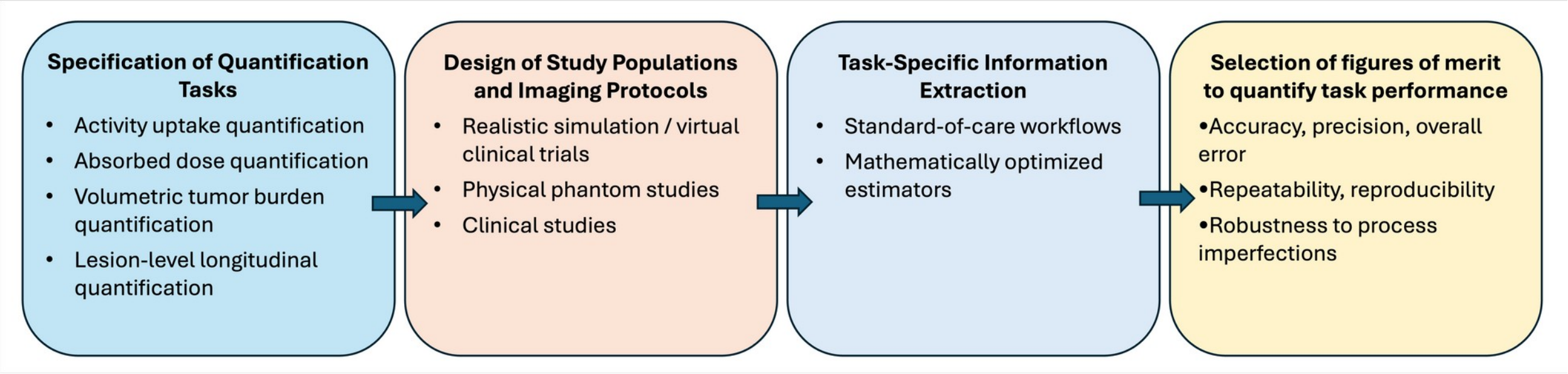

Figure 2. Four design steps in the EQUIP-RPT framework for objective, task-based evaluation of quantitative imaging methods in RPT.

Together, these steps define the study elements combined into an explicit evaluation claim: a statement of the QI method performance for a prespecified quantification task in RPT under specified study conditions.

### 2.1 Quantification Tasks

This step specifies the quantification tasks that imaging methods are intended to support. Below, we introduce four foundational tasks that provide direct inputs to therapy planning, monitoring, and safety assessment, as illustrated in Figure 3.

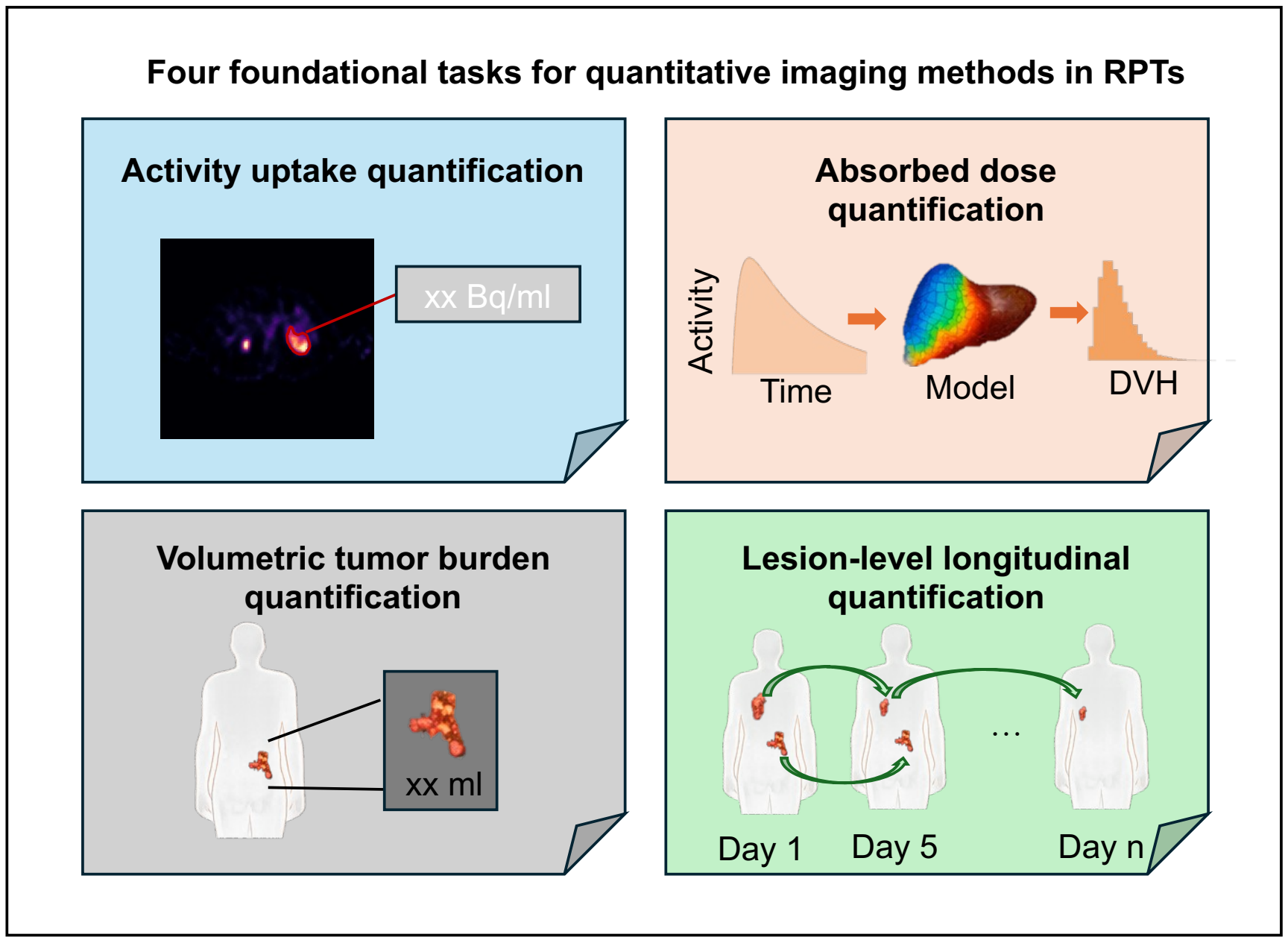

Figure 3. Four foundational tasks for quantitative imaging methods in RPTs that provide direct inputs to therapy planning, monitoring, and safety assessment. DVH, dose-volume histogram.

### 2.1.1 Activity Uptake Quantification

This task is to estimate the spatial distribution of radiopharmaceutical uptake and report activity (Bq) or activity concentration (Bq/mL) either voxel-wise or within defined volumes of interest (VOIs) at specific post-injection time points. Although these estimates often serve as inputs to downstream tasks such as absorbed-dose quantification, activity uptake quantification is treated here as distinct because the uptake distribution may itself be the target of estimation, supporting biodistribution evaluations, treatment planning, and uptake-based biomarker development. For example, for radium-223-based RPT, the task can be quantifying activity uptake of osteoblastic bone metastases with intestinal VOIs included. In lutetium-177-DOTATATE therapy, the task can be quantifying uptake of somatostatin receptor–positive lesions and organs at risk such as kidneys and liver.

### 2.1.2 Absorbed Dose Quantification

The task is to quantify absorbed dose, defined as the energy imparted per unit mass, in gray (Gy), for lesions and radiosensitive organs. In internal dosimetry, the linkage from quantified activity uptake to absorbed dose is the time-integrated activity (TIA; MBq·h), typically obtained by fitting and integrating time–activity curves[11]. Methods have also been developed to estimate absorbed dose from single-time-point measurements and to perform predictive dosimetry to support therapy planning or adaptation[4].

Dose is reported at the VOI, voxel, or finer spatial scale according to intended use. VOI-level dose remains common for many therapies. In lutetium-177-based RPTs, the mean organ and lesion doses are derived from VOI TIAs using established formalisms[26]. When spatial heterogeneity is clinically relevant, voxel-level dosimetry provides three-dimensional dose maps and dose–volume histograms (DVHs) that complement VOI means[16]. For alpha-emitters, the short particle range and high linear energy transfer motivate evaluation at finer spatial scales; micro- or sub-voxel–level quantifications have been explored when voxel averages may not reflect biologically relevant energy deposition[27].

### 2.1.3 Volumetric Tumor Burden Quantification

This task quantifies tracer-defined volumetric tumor burden. The tumor burden may be quantified by volume, defined as the sum of tracer-avid lesion volumes, or as lesion volume multiplied by mean uptake, under a prespecified segmentation rule[28]. The reporting levels commonly include whole-body, organ-level, and lesion-level summaries.

In prostate-specific membrane antigen (PSMA)-expressing prostate cancer, commonly used measures include whole-body total tumor volume (PSMA-TV) and the uptake-weight volume, such as total lesion PSMA uptake[29]. In neuroendocrine tumors imaged with somatostatin receptor (SSTR) PET, analogous estimates include somatostatin receptor-expressing tumor volume (SRETV) and total lesion SSTR uptake[30]. These volumetric tumor burden measures are primarily used for baseline risk stratification and response assessment.

### 2.1.4 Lesion-Level Longitudinal Quantification

This task establishes lesion correspondence across serial imaging time points and quantifies lesion-specific estimates such as uptake, time-integrated activity, absorbed dose, or volumetric change, supporting individualized response assessment and dosimetry. Lesion-level assessment is an increasingly important component of RPT. Current practice, however, often reduces heterogeneous disease patterns to patient-level scores or a limited set of target lesions, overlooking substantial inter-lesion variability in treatment response[31]. This is particularly problematic in high-burden diseases such as metastatic castration-resistant prostate cancer and neuroendocrine tumors, where patients may present with tens to hundreds of metastases[32].

Lesion tracking provides the mapping required for longitudinal quantification[32,33]. Automated or standardized tracking methods are therefore important: without them, analyses are often restricted to a small subset of manually selected lesions, limiting scalability and potentially biasing quantifications. Tracking methods have been developed for computed tomography (CT), magnetic resonance imaging (MR), PET, and PET/MR, using deformable registration, graph-based matching, and AI-driven approaches, but fewer efforts have targeted SPECT, despite its central role in post-therapy quantification for RPT[32,33].

Growing evidence supports the clinical relevance of lesion-level longitudinal quantification. PSMA-PET-derived tumor-burden measures are prognostic in prostate cancer, and the Standardised PSMA PET/CT Analysis and Reporting Consensus (SPARC) supports standardized reporting of PSMA volume alongside uptake measures[34]. Similarly, early longitudinal SPECT-based tumor-burden measures after lutetium-177-PSMA therapy have shown associations with progression-free and overall survival[35,36]. These developments underscore lesion-level longitudinal analysis as a distinct quantification task within RPT workflows, alongside traditional activity, absorbed dose, and volumetric tumor burden estimation tasks.

### 2.2 Study Populations and Imaging Protocols

For task-based evaluation studies to be clinically relevant, it is important that the study population be clinically representative and the imaging process be similar to clinical protocols. The study population, including clinical patients, physical phantoms, or virtual patients, should ideally capture the sources of variability that are relevant to the target application, including demographics, disease biology, tumor burden and spatial distribution, organ function, administered activity, and imaging schedule. A best practice is prospectively defining inclusion and exclusion criteria, sampling strategy, number of sites, and any prespecified stratification factors.

The imaging protocol should ideally be specified with comparable rigor, including modality, scanner configuration, acquisition timing and duration, relevant acquisition parameters, and the full correction, calibration, and reconstruction pipeline with associated software versions, when applicable, because quantitative performance can vary significantly with these choices. Section 3 describes how these general design principles are implemented in simulation, physical phantom, and clinical studies.

### 2.3 Task-Specific Information Extraction

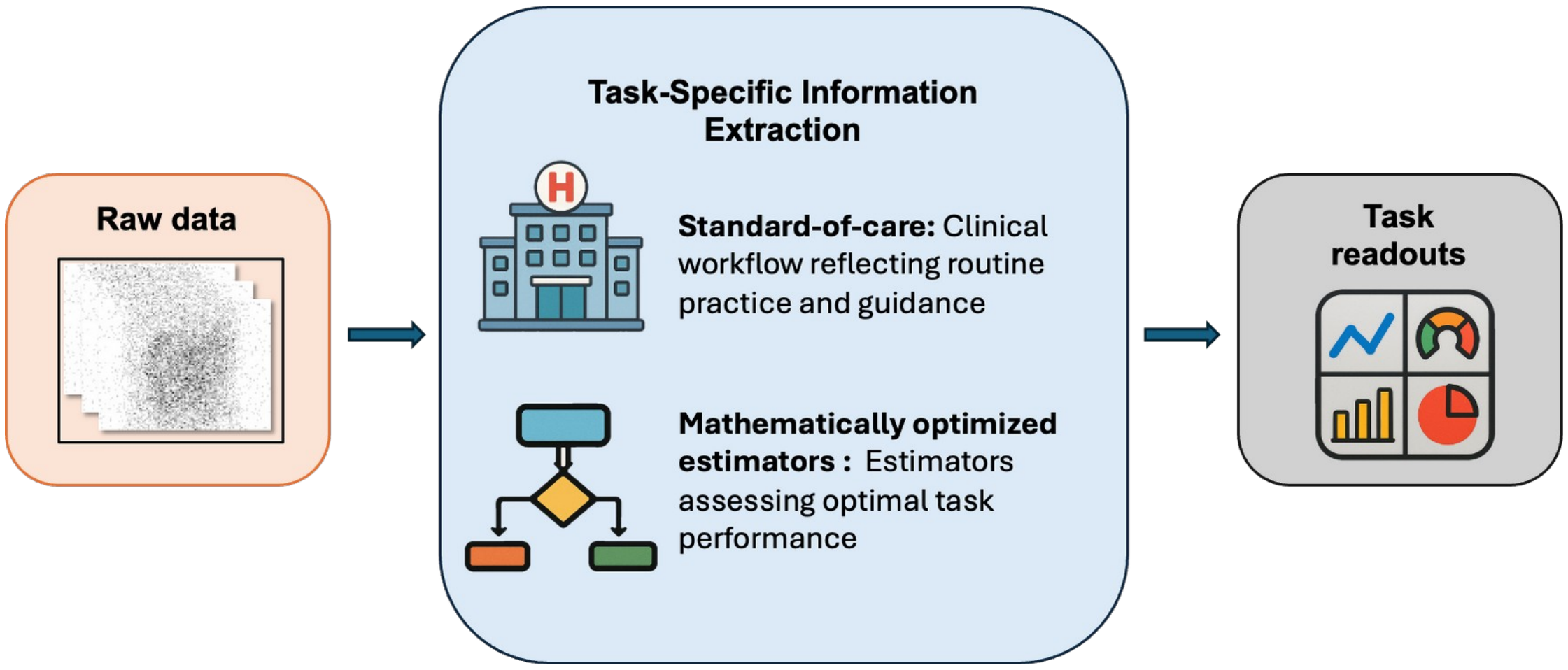

Figure 4. Two paradigms for task-specific information extraction: standard-of-care and mathematically optimized estimators.

Task-specific information extraction refers to the analysis pipeline that converts acquired data into estimates corresponding to the prespecified clinical task. To control confounding, the quantification protocol is best designed and specified prospectively. This prospective specification is particularly important for tasks like longitudinal and lesion-level quantification, where small inconsistencies can propagate into substantial differences in quantitative estimates[15,32].

Figure 4 illustrates two paradigms for task-specific information extraction. Standard-of-care workflows are appropriate when the goal is to evaluate performance under current clinical practice. In this case, the reconstruction and correction pipeline, calibration procedures, VOI definitions, partial-volume correction rules, normalization or harmonization procedures, segmentation, registration, lesion tracking methods, and software versions are best specified to reflect routine clinical practice and relevant guidance[11,28]. When comparing QI methods, these pipeline components are best matched where appropriate and optimized for each method as needed, using the same prespecified task-performance criterion. For example, differences in SPECT count levels can affect the trade-off between bias and variance, and thus the number of reconstruction iterations appropriate for each method. The iteration number may be optimized for each QI method using the same prespecified task-performance criterion, such as minimizing the mean squared error of the estimated VOI-mean activity uptake.

In contrast, mathematically optimized estimators are appropriate when the goal is to assess the task-relevant information available from the QI method under specified statistical assumptions [19]. By accounting for the statistics of each QI method, this paradigm reduces confounding arising from a mismatch between the information-extraction process and those statistics. It therefore provides a complementary assessment of the performance achievable when information extraction is tailored to the QI method rather than constrained by an existing clinical workflow. Mathematically optimized estimators can be broadly categorized into frequentist and Bayesian approaches. Frequentist estimators treat parameters as fixed but unknown quantities. A widely used frequentist estimator is Maximum Likelihood Estimation (MLE), well-known for being asymptotically unbiased and achieving the lowest possible variance for any unbiased estimator, as given by the Cramér–Rao lower bound, provided such an efficient estimator exists[19]. Bayesian methods treat parameters as random variables following assumed known prior distributions. Bayesian estimators determine parameter estimates by minimizing a defined cost function. For instance, the posterior mean estimator minimizes ensemble mean square error (EMSE). The Maximum a Posteriori (MAP) estimator minimizes a cost function sharply focused on closely matching the true parameter value.

When optimal estimators of unrestricted form are unavailable or computationally impractical, optimal linear estimators can be useful. For example, Wiener estimators minimize EMSE within the class of linear estimators and offer computational advantages by requiring only the relevant first- and second-order statistics. They enable objective comparisons of QI methods considering the best achievable performance within the linear estimator class. Related approaches, such as scanning linear estimation, have also shown promising performance in quantification tasks and have been used in SPECT system and protocol optimizations[37]. Recent advancements, including the Wiener INtegration Projection-Domain Quantification (WIN-PDQ) method, have integrated

Wiener estimation into projection-domain quantification while accounting for heterogeneous uptake distributions within VOIs[38].

The optimality of these estimators is established mathematically for specified performance criteria and estimator classes under assumed statistical models. However, the numerical implementation needs to be validated before using mathematically optimized estimators for task-specific information extraction. For example, the numerical implementation can be checked using independent and, when necessary, simplified numerical experiments, where the estimator output can be obtained analytically and thus provides a benchmark for comparison. Although many such estimators, including projection-domain approaches, are not yet standard-of-care, they can become clinically applicable as the evidence base matures.

### 2.4 Selection of Figures of Merit to Quantify Task Performance

Figures of merit (FoMs) are quantitative metrics to characterize the performance of a QI method. They are best selected according to the evaluation objective, study design, and ground-truth availability. In RPT, key attributes include accuracy, precision, overall error, repeatability, and reproducibility[39]. For longitudinal or multi-visit tasks, robustness to clinically plausible protocol deviations is also important. Details and mathematical definitions of the FoMs summarized below are provided in Appendix A.

When ground truth is known, as in many simulation and physical phantom studies, FoMs can directly characterize estimation performance. Accuracy describes systematic deviation of the estimates from the ground truth and is commonly measured by bias. Estimator precision describes variability of estimates across repeated noise realizations for a fixed true parameter and is often measured by variance, standard deviation, or coefficient of variation. Overall error combines accuracy and precision and is commonly measured by mean squared error (MSE) or root MSE (RMSE). For population-level evaluation, ensemble MSE (EMSE) or root EMSE (REMSE) is often appropriate[25]. These ground truth-based FoMs are especially relevant when absolute accuracy is needed, for example, for absolute dose quantifications.

When the evaluation objective is stability, agreement, or comparability, repeatability and reproducibility become important. Repeatability quantifies agreement under nominally identical conditions, such as test–retest imaging with a fixed protocol and analysis workflow, and can be summarized by Bland–Altman limits of agreement or the within-subject coefficient of variation[40]. Reproducibility quantifies agreement across scanners, sites, operators, or readers, and the intraclass correlation coefficient (ICC) is commonly used. Because multiple ICC formulations exist, the selected model, type, definition, estimation method, and confidence interval should be reported[41].

For longitudinal workflows, an additional performance attribute is the robustness to clinically plausible deviations from a prespecified acquisition or analysis process, such as changes in sampling schedule or scan duration[42]. Robustness can be quantified by the relative degradation in an overall error metric, such as EMSE.

In most patient-data evaluations, direct ground truth is unavailable. In this case, the FoM selection can typically place greater emphasis on repeatability, reproducibility, and robustness. Another viable alternative is to use reference standards. However, when the errors in the reference standard are high, the evaluation may be inaccurate[43]. Thus, reference standards need careful selection and cautious interpretation. Towards addressing the issue of lack of ground truth, approaches such as no-gold-standard evaluation (NGSE) techniques have been proposed[44–46]. These techniques are built on the premise that the true and measured quantitative values are statistically related and are designed to estimate the parameters that describe the relationship without knowledge of the true value. The techniques have demonstrated strong promise in evaluating QI methods, including quantitative SPECT methods for RPTs[44,47].

### 2.5 Formulating the Evaluation Claim

The outputs of the four design steps form an explicit evaluation claim. Here, we use the term evaluation claim to denote a specific statement of the performance of the QI method under evaluation for a prespecified quantification task in RPT, under the study conditions specified by the evaluation design[25].

The claim ideally specifies the quantification task, the population used in the evaluation, including clinical patients, virtual patients, or physical phantoms, the imaging process, including acquisition and reconstruction approaches, the task-specific information-extraction process, and the FoMs used to characterize the task performance[25]. When relevant, the claim also states the basis for comparison, such as a reference standard or a comparator method, together with the key assumptions and operating conditions under which the claim is intended to hold. For claims of improvement over existing methods, comparators should include relevant standard-of-care or well-established approaches for the intended clinical task, with the choice justified in relation to the claim. Comparator methods should be optimized as needed to avoid favoring the proposed method through weak comparators or suboptimal settings.

## 3. Implementation of EQUIP-RPT Framework

The four-step framework above defines the elements used to specify an objective task-based evaluation of QI methods in RPTs. Their implementation, however, differs across simulation, physical phantom, and clinical studies because these settings provide different levels of control, access to ground truth, and clinical realism. As summarized in Table 1, these studies have different tradeoffs[25].

Table 1. Comparison of evaluation settings for objective, task-based evaluation of quantitative imaging methods in radiopharmaceutical therapy.

| Evaluation setting | Cost | Experimental control | Realism | Main note |
|---|---|---|---|---|
| **Simulation study** | Low | High | Medium | Known ground truth; flexible; model quality is important |
| **Physical phantom study** | Medium | Medium–High | Medium–High | Known filled activity; applicable in calibration and harmonization; limited biological realism |
| **Clinical study** | High | Low–Medium | Highest | Most clinically translatable; often lacks direct ground truth; constrained by ethics and logistics |

### 3.1 Simulation Studies

Simulation studies are most useful during early stages of method development, when the goals are to characterize performance over a broad range of patient, lesion, and protocol setups, as well as to inform the design and statistical power of subsequent phantom or clinical studies.

Realistic simulation studies for RPTs, particularly VITs, enable evaluation over representative virtual populations with known ground truth, while allowing systematic variation of acquisition, reconstruction, and analysis settings at low cost[22]. These studies require both realistic objects and imaging models. On the object side, such studies commonly employ anthropomorphic digital phantoms such as Extended Cardiac-Torso (XCAT), which support controlled variation of anatomy and physiological motion[48]. Parameters used to simulate the virtual patient population can be derived from published demographic data or clinical records. For RPTs, lesion modeling is particularly important. Lesion number, volume, spatial distribution, and uptake can be derived from clinical images or records[49]. When relevant, lesion shape and intra-lesion uptake heterogeneity can also be modeled from clinical data to improve clinical realism, since these factors can affect quantification performance[50]. Approaches that insert simulated signals into actual clinical images or projection data can help improve the background realism[51–53].

On the imaging side, PET or SPECT simulations should ideally model all relevant image-degrading processes. For SPECT, this typically includes collimator geometry and materials, detector response, scatter, attenuation, and isotope-specific emissions. For PET, it typically includes scanner geometry and timing properties, attenuation, randoms, scatter, and spatial-resolution-degrading effects such as positron range, non-collinearity, and inter-crystal penetration. Representative simulation tools are summarized in Appendix Table 1. For RPT applications, it is recommended to validate SPECT simulations for the specific isotope under consideration, because many RPT isotopes have complex gamma-ray emission spectra with high-energy peaks that may not be fully covered by validations developed for conventional SPECT simulations. In addition, for clinically relevant claims, the simulated scanning protocol should reflect clinical practice.

### 3.2 Physical Phantom Studies

Physical phantom studies provide a bridge between simulations and clinical evaluations of QI methods for RPTs. They support direct evaluation of bias, variance, and reproducibility and are also valuable for validating simulation workflows and for calibration and harmonization across systems, thereby improving consistency in clinical evaluations[54].

Phantom choice is guided by the quantitative task and the evaluation objective. Appendix Table 2 summarizes representative phantom types and their primary uses. Simple geometric phantoms are useful for benchmarking activity estimation and assessing partial-volume correction under controlled conditions. Anthropomorphic or organ-specific phantoms, including 3D-printed or fillable models of the torso, kidney, or liver, provide more realistic attenuation and scatter conditions and are better suited to evaluation of activity or activity-concentration estimation and VOI-based quantification[55–58]. Dynamic or multi-compartment phantoms can emulate biological kinetics and are therefore useful for evaluating time-integrated activity estimation and temporal sampling protocols[59], whereas heterogeneous phantoms can emulate non-uniform lesion uptake and tissue heterogeneity[60]. For phantom studies intended to be clinically relevant, phantom design parameters should reflect clinically relevant activity concentrations, contrast levels, and geometry for the radiopharmaceutical and task of interest.

For SPECT in RPT, isotopes may have unique imaging characteristics, such as complex emission spectra, so physical-phantom study should be designed accordingly. For example, these isotopes often have multiple photopeaks, thus motivating multi-energy-window acquisition. When available, the acquisition protocol can follow common clinical practice or established recommendations[8], else a best practice is optimizing the collimator and the energy-window settings specific to the isotope. Before acquisition, peaking the camera for all selected windows is important, since peak shifts can cause count loss and bias quantification. Absolute quantification requires sensitivity calibration factors that convert the reconstructed values to activity or activity concentration. The calibration factor is best established and verified for each scanner, isotope, energy window, and reconstruction configuration[8,14]. Dead-time- and pile-up-induced count losses may be non-negligible for radionuclides such as lead-212, and these effects are often specific to the radionuclide, count rate, and system configuration[61]. Compensation for these count losses needs to be derived and validated. Scanning protocols for NaI- and CZT-based SPECT scanners can differ substantially and should therefore be established separately. For longer-lived isotopes, residual activity from previous fills should be removed or allowed to decay before phantom reuse.

In multicenter settings, identical phantom fills and standardized analysis templates are important. Harmonization programs such as EANM EARL[62] and the SNMMI Clinical Trials Network scanner validation[63] for PET, together with emerging initiatives for quantitative SPECT/CT[7,14], demonstrate that use of common phantoms and shared protocols can substantially reduce inter-site variability. Adopting a similar approach for evaluations of QI methods in RPTs, based on traceable activity standards and shared reconstruction protocols, can facilitate comparisons across centers.

### 3.3 Clinical Studies

Clinical studies provide the strongest clinical evidence for evaluating QI methods in RPTs because they evaluate performance with patient populations, capturing aspects such as physiological variability and operational factors that can be difficult to reproduce in simulations or phantom studies and can support continued optimization of acquisition protocols and analysis pipelines during translation to routine use. However, these studies are resource-intensive and frequently limited by the absence of ground truth.

Clinical study design needs to balance practical constraints, statistical power, and ethical feasibility. Study cohorts are typically aligned with the intended-use context and may represent the relevant clinical variability. Inclusion and exclusion criteria, plans for calibrations, harmonization, and reproducibility assessments are typically specified prospectively. Multi-time-point imaging and quantification protocols should generally be justified and, where possible, validated by prior simulation or phantom studies[3,11,64]. Because ground truth is generally unavailable, clinical evaluations commonly rely on reference or consensus workflows for comparison[11,28], and acquisition protocols can follow standardized procedures established through phantom studies, with cross-calibration and harmonization applied to reduce inter-site variability.

Ethical and operational planning is also important, including justification of any imaging beyond standard care, adherence to the as-low-as-reasonably-achievable principle, and governance for anonymization, data sharing, and centralized analysis in multicenter studies. Large-scale multicenter studies remain particularly important for establishing generalizability, and hybrid validation strategies that integrate simulation or phantom evidence with clinical measurements can strengthen interpretation and support regulatory and clinical acceptance[54,65].

### 3.4 Example Application of the EQUIP-RPT Framework

To illustrate the application of EQUIP-RPT framework in practice, consider a case study with the goal of early-stage evaluation of SPECT systems for alpha-particle RPT on the task of estimation of VOI-mean activity concentration (Bq/mL) in prospectively defined organ and lesion VOIs at fixed post-injection time points. Given the goal of early-stage evaluation, we can consider a VIT allowing direct estimation of bias, variance, and overall error. The virtual cohort could be based on XCAT[48], with realistic anatomical and clinically derived physiological variability, lesion burden, uptake patterns, and intra-regional heterogeneity. The imaging model could be implemented using Monte Carlo SPECT simulation software such as SIMIND that has been validated using physical-phantom studies[66–68]. Task-specific information could be extracted using a posterior-mean estimator, which minimizes EMSE under the specified imaging and population models[19]. The primary FoM is REMSE of uptake estimates across the virtual population. Ensemble bias and variance are reported separately, and uncertainty could be quantified using bootstrap confidence intervals across cases and noise realizations. Figure 5 summarizes the instantiation of the four design steps in this example, and Appendix D provides implementation details.

The corresponding evaluation claim could be as follows: *In a virtual imaging trial, under the prespecified acquisition conditions, the evaluated SPECT system yielded lower REMSE than the comparator system for estimation of VOI-mean activity concentration (Bq/mL) in prespecified organ and lesion VOIs at fixed post-injection time points.*

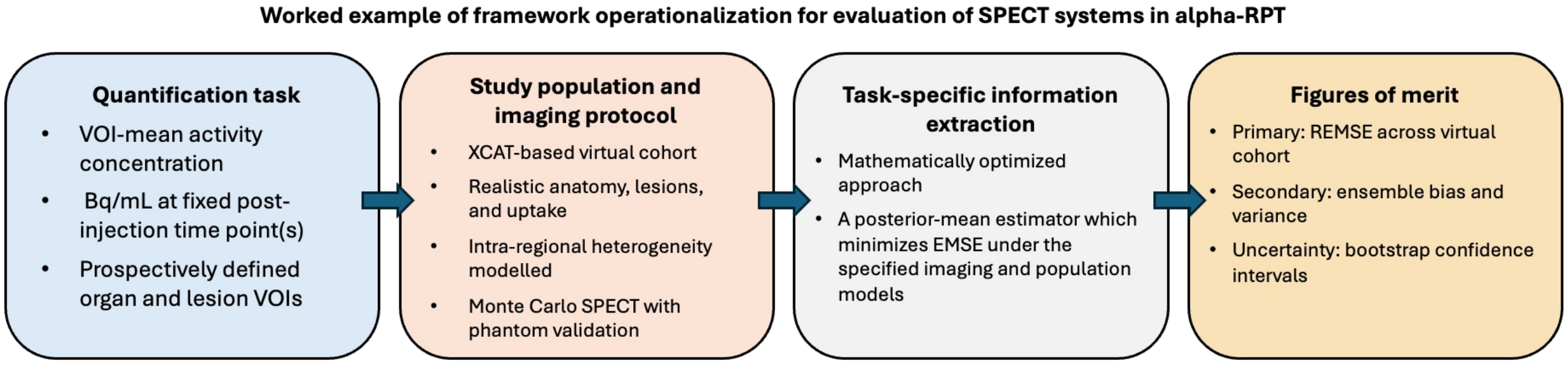

Figure 5. Example application of the EQUIP-RPT framework for evaluating a SPECT system developed for alpha-RPTs.

## 4. Discussion and Summary

This work outlines the Evaluation of Quantitative Imaging Performance in Radiopharmaceutical Therapy (EQUIP-RPT) framework for objective task-based evaluation of QI methods in RPTs. We organize evaluation around four linked design elements: (i) specification of the quantification task, (ii) design of the study population and imaging protocol, (iii) task-specific information extraction, and (iv) selection of figures of merit (FoMs). We describe how evaluations can be implemented across simulation, physical phantom, and clinical studies. By structuring evaluation around common design considerations while accounting for setting-specific constraints, EQUIP-RPT is intended to make evaluations more comparable, clinically interpretable, and robust to confounding by uncontrolled variation in study design.

We confined the scope of this work to direct quantification tasks in RPT because their outputs provide foundational inputs to treatment planning, monitoring, and safety assessment. Evaluation of QI methods for predicting clinical outcomes was not included, since such tasks require integration with radiobiological, mechanistic, pharmacokinetic, and biostatistical models beyond the scope of the present framework. These downstream outcome prediction and treatment optimization efforts are closely related to the emerging concept of computational nuclear oncology (CNO), which aims to integrate quantitative imaging, dosimetry, modeling, artificial intelligence, and digital-twin approaches to support precision RPT[69,70]. While the proposed framework is not intended to evaluate full CNO pipelines, it strengthens a prerequisite for them by specifying how QI methods can be objectively evaluated on foundational quantitative tasks before these estimates are used in outcome prediction or treatment adaptation models.

Key directions to strengthen this framework include the development and validation of rigorous estimation algorithms for quantification tasks. Analogous to the optimal observer used for detection, optimal estimators can provide important mechanisms for extracting task-specific information. Phantoms with known activity concentrations can support estimator validation by providing ground truth for characterizing bias, variance, and overall error under controlled conditions. Developing and validating such estimators and associated quantification task-based image-quality measures, and integrating them into practical evaluation studies, is thus an important research area. Progress along this line would enable more objective, reproducible, and theoretically grounded assessment of QI methods in RPT.

In parallel, the development of advanced anthropomorphic phantoms remains an active area of research. This includes 3D-printed and tissue-equivalent designs that support dynamic filling, realistic lesion heterogeneity, and other features relevant to RPT studies. Additionally, as scanner spatial resolution improves, traditional phantom designs may not always provide sufficient structural detail to adequately challenge modern systems, further motivating the development of high-resolution physical phantoms for quantitative validation. Integration of phantoms with automated filling systems and activity-traceable validation pipelines could further improve reproducibility and validity. Future efforts emphasizing standardized phantom datasets and shared protocols for RPT dosimetry can strengthen the role of phantom studies in task-based evaluation and multicenter standardization.

Overall, EQUIP-RPT provides a task-based framework for objective evaluation of QI methods in RPT across simulation, phantom, and clinical settings, and shows how these components can be combined into an explicit evaluation claim. The aim is not to prescribe a single path, but to provide a coherent set of study-design options that support rigorous evaluation and, ultimately, more reliable clinical adoption of QI in RPT.

## Appendix A: Mathematical Definitions and Details of Figures of Merit

Denote $\theta$ as the true task-specific parameter to be quantified, and denote $g$ as the acquired imaging data. For simplicity in notation, we define $\theta$ a scalar. Multicomponent estimates can be handled component-wise. An estimation approach $T$ produces an estimate

$$\hat{\theta} = T(g).$$

For a fixed $\theta$, the data $g$ are random due to acquisition noise and other stochastic effects. The conditional distribution of $g$ given $\theta$ is denoted by $Pr(g/\theta)$. In addition, $\theta$ varies across the target population according to a distribution denoted by $Pr(\theta)$.

### A.1 Accuracy

Bias is defined as

$$b(\theta) \triangleq E_{g/\theta}[\hat{\theta}] - \theta,$$

where $E_{g/\theta}[\cdot]$ denotes expectation with respect to $Pr(g/\theta)$. Since bias typically depends on $\theta$, bias profiles $b(\theta)$ across clinically relevant ranges are generally more informative than a single scalar summary.

When a population-level summary is preferred, the ensemble bias is defined as

$$\bar{b} \triangleq E_{\theta}[b(\theta)],$$

where $E_{\theta}[\cdot]$ denotes expectation with respect to $Pr(\theta)$.

### A.2 Estimator precision

Estimator precision is typically quantified by the variance of $\hat{\theta}$ across repeated noise realizations for a fixed $\theta$. Define

$$v(\theta) \triangleq E_{g/\theta}\left[\left\|\hat{\theta} - E_{g/\theta}[\hat{\theta}]\right\|_2^2\right],$$

and the corresponding standard deviation

$$\sigma(\theta) \triangleq \sqrt{v(\theta)}.$$

For scale-normalized reporting, the coefficient of variation (CV) is defined as

$$CV(\theta) \triangleq \frac{\sigma(\theta)}{|E_{g/\theta}[\hat{\theta}]|}.$$

A population-level summary is given by the ensemble variance

$$\bar{v} \triangleq E_{\theta}[v(\theta)].$$

### A.3 Overall Error

Mean squared error (MSE) is defined as

$$MSE(\theta) \triangleq E_{g/\theta}\left[\left\|\hat{\theta} - \theta\right\|_2^2\right].$$

A population-level reliability summary is the ensemble mean squared error (EMSE):

$$EMSE \triangleq E_{\theta}\left[MSE(\theta)\right].$$

For reporting in the original units of the targeted estimate, the corresponding root measures are reported as the root mean squared error (RMSE) and the root ensemble mean squared error (REMSE), defined by

$$RMSE(\theta) \triangleq \sqrt{MSE(\theta)}, REMSE \triangleq \sqrt{EMSE}.$$

**A.4 Repeatability**

Repeatability quantifies agreement when imaging is repeated under identical conditions. A common design is a test–retest study in which each subject or phantom is imaged twice within a short interval so that the underlying true parameter can be treated as unchanged. Denote $x_{i1}$ and $x_{i2}$ as the two imaging-derived task readouts for subject $i$, where $i=1,\ldots,N$.

A standard summary is based on paired differences

$$d_i \triangleq x_{i2} - x_{i1}, i=1,\ldots,N.$$

The mean difference

$$\bar{d} \triangleq \frac{1}{N}\sum_{i=1}^{N} d_i$$

quantifies systematic offset between test and retest, while the standard deviation of the differences

$$s_d \triangleq \sqrt{\frac{1}{N-1}\sum_{i=1}^{N}\left(d_i - \bar{d}\right)^2}$$

quantifies the dispersion. Bland–Altman limits of agreement are given by

$$\bar{d} \pm 1.96\, s_d,$$

which provides an interpretable interval for the expected test–retest difference under an approximate normality assumption for the differences.

A commonly used dispersion measure is the within-subject standard deviation. For two repeats,

$$s_w \triangleq \frac{s_d}{\sqrt{2}}.$$

A scale-normalized repeatability metric is the within-subject coefficient of variation (wCV):

$$wCV \triangleq \frac{s_w}{\bar{x}}, where\ \ \bar{x} \triangleq \frac{1}{2N}\sum_{i=1}^{N}\left(x_{i1} + x_{i2}\right).$$

## A.5 Reproducibility

Reproducibility quantifies agreement of the estimates when imaging conditions vary, such as changes in scanner, site, operator, or analysis team. A widely used summary is the intraclass correlation coefficient (ICC). Following Koo and Li[41], the intraclass correlation coefficient is not a single statistic. The reported intraclass correlation coefficient should be matched to the study design and should explicitly specify the model, the type, and the definition. The model is selected from one-way random effects, two-way random effects, or two-way mixed effects. The type specifies whether the statistic corresponds to a single imaging acquisition or the mean of $k$ acquisitions or raters. The definition specifies whether the target is absolute agreement or consistency. The estimation method or software and the 95% confidence interval should also be reported. For interpretation, intraclass correlation coefficient values below 0.50, between 0.50 and 0.75, between 0.75 and 0.90, and above 0.90 are commonly described as indicating poor, moderate, good, and excellent reliability, respectively[41].

## A.6 Robustness to Process Imperfections

This category of FoMs quantifies the stability of task performance under clinically plausible deviations from a reference acquisition and analysis process. Relevant deviations include changes in the imaging schedule, the number or timing of sampling time points, scan duration, and prespecified analysis choices[42,64,71,72]. One practical robustness summary is the relative degradation in overall error, assessed by comparing an overall error metric such as EMSE under the perturbed process to that under the reference process. The robustness to process imperfections can also be characterized through sensitivity analyses that quantify how bias and variance change as the process departs from the reference. For continuous perturbations, local sensitivity can be reported using changes in bias and variance per unit perturbation around the reference. For discrete perturbations, such as removal of time points or temporal shifting of acquisitions, similar finite-difference summaries can be used. These robustness summaries should be evaluated over clinically relevant ranges of the task parameter and over perturbations that reflect potential variability in practice. For decision-threshold tasks, robustness to process imperfections can be characterized as the probability that the task readout, or a derived estimate such as time-integrated activity, absorbed dose, or change in tumor burden, remains within a prespecified clinically acceptable tolerance under missing or delayed acquisitions.

## Appendix B: Representative Simulation Tools

Appendix Table 1 summarizes representative simulation tools for PET and SPECT, including their modeling approaches, modeled physical effects, and computational characteristics.

**Appendix Table 1.** Representative simulation tools developed for modeling PET and SPECT imaging systems.

| APPROACH | TOOL | MODALITY | EFFECTS MODELED | FEATURE AND SPEED |
|---|---|---|---|---|
| Monte Carlo | GATE[73] | PET & SPECT | Particle-level transport including Compton and photoelectric effects; optional optical-photon tracking; detailed detector and timing modeling | General-purpose Geant4-based Monte Carlo with high physical fidelity; typical whole-scan runtimes span hours to days on central processing units (CPUs) (slow) |
| | SimSET[74] | PET & SPECT | Photon creation and transport in voxelized objects with modular collimator and detector modeling | Emission-tomography Monte Carlo designed for efficiency; typically an order of magnitude faster than Geant4-based tools (fast) |
| | PeneloPET[75] | PET | PET simulations using PENELOPE physics, including positron range and detector response | PET-dedicated Monte Carlo with improved transport efficiency but still limited by photon statistics (medium) |

| APPROACH | TOOL | MODALITY | EFFECTS MODELED | FEATURE AND SPEED |
|---|---|---|---|---|
| | SIMIND[66–68] | SPECT | SPECT camera modeling including collimator and detector response, scatter, and septal penetration; validated against phantom measurements | Established SPECT Monte Carlo; parallelizable with moderate computation time depending on photon statistics (medium) |
| Analytical model-based | ASIM[76] | PET | Deterministic forward projection with attenuation, scatter and randoms approximations, and point-spread modeling | Analytic simulator designed for rapid generation of sinograms and images compared with Monte Carlo (fast) |
| | SMART-PET[77] | PET | Analytic forward model with attenuation and blur; configurable scatter and randoms; optional time-of-flight modeling | Reported runtime about 3–15 minutes per simulated image (fast) |
| | PETSTEP[78] | PET | Modeling scanner blur and attenuation; approximate scatter handling | Reported runtime about 4–6 minutes per image on a 2.7 GHz CPU (fast) |

| APPROACH | TOOL | MODALITY | EFFECTS MODELED | FEATURE AND SPEED |
|---|---|---|---|---|
| | FAST-PET[79] | PET | Analytical projector with scatter and randoms estimation validated against GATE Monte Carlo | Achieves roughly 100-fold lower computation time than GATE with comparable quantitative accuracy (fast) |
| | STIR[80] | PET & SPECT | Analytical projectors and utilities for simulating projections or sinograms; widely used reconstruction framework | Provides fast analytic projections; runtime depends on chosen projector implementation (fast) |

## Appendix C: Representative Physical Phantoms

Appendix Table 2 summarizes representative physical phantom types, their primary applications, advantages, and limitations for quantitative imaging evaluation.

**Appendix Table 2.** A concise summary of representative phantom types, primary uses, key advantages, and limitations.

| PHANTOM TYPE | EXAMPLE | PRIMARY USE | ADVANTAGES | LIMITATIONS |
|---|---|---|---|---|
| Geometric | NEMA IEC[81], Jaszczak[82] Q3P[57] | Activity recovery Calibration | Simple, reproducible | Limited anatomical realism |
| Organ-specific | 3D-printed organ inserts[55,83] | Organ-level activity quantification | Realistic attenuation and uptake | Custom fabrication needed |
| Anthropomorphic | Probe-IQ[58] | Patient mimicking for PET and SPECT | High clinical realism | Costly, limited motion |
| Dynamic | Modular multi-compartment[84] | Kinetic or time-activity studies | Non-static; suitable for kinetic modeling | Complex setup |
| Multi-center standardized | NEMA IQ phantom[62] and CTN3/CTN5 anthropomorphic chest phantom[63,85] | Cross-site harmonization, clinical trial validation | Proven reproducibility | Fixed geometry |

## Appendix D: Example Application Details

This section illustrates one possible implementation of the proposed framework for objective, task-based evaluation of SPECT systems in alpha-particle radiopharmaceutical therapy. It is intended as a worked example rather than a prescriptive study design.

### D.1 Specification of Quantification Task

The target task is the estimation of VOI-mean activity concentration (Bq/mL) at fixed post-injection time points in prospectively defined organ and lesion VOIs.

### D.2 Design of Study Population and Imaging Protocol

Because the objective is to compare absolute quantitative performance across systems, the example uses a virtual clinical trial, in which the underlying activity distributions are known by construction. An anthropomorphic XCAT-based cohort is used to model clinically relevant variability in body habitus, organ size, lesion number, lesion location, lesion morphology, and regional uptake. Intra-regional uptake heterogeneity is modeled in both organs and lesions. All study factors other than the SPECT systems under comparison are selected to replicate those in typical clinical conditions and are matched across the systems being compared.

SPECT acquisition is modeled using Monte Carlo simulations, including relevant emission spectra, attenuation, scatter, and full collimator-detector response. Targeted validation against physical phantom measurements is used to support model fidelity. The study design also prespecifies the number of virtual patients, inclusion and exclusion criteria, and the number of noise realizations per condition. Random seeds and software versions are recorded to support reproducibility.

### D.3 Task-Specific Information Extraction

Task-specific information extraction is performed using a prespecified mathematically optimized estimator so that system comparisons reflect task-relevant information content rather than being confounded by the suboptimal downstream analysis process. In this illustrative example, VOI-mean activity concentrations would be estimated using their posterior means given the projection data. For each system, the estimator would use the corresponding data likelihood and a common joint prior describing regional uptake and intra-regional heterogeneity, consistent with the population model in Appendix D.2. Heterogeneity parameters would be marginalized when estimating VOI-mean activity concentrations. Under this model, the posterior-mean estimator minimizes EMSE, and hence REMSE, among all estimators based on the projection data[19]. It therefore provides an objective approach for comparing system performance on the specified quantification task.

### D.4 Figures of Merit

Because the uptake distribution is known by construction, bias, variance, and overall error can be estimated directly. The primary FoM is the REMSE of VOI-mean uptake estimates across the virtual population. Ensemble bias and variance are reported separately as secondary FoMs. Uncertainty in all reported FoMs is quantified using bootstrap confidence intervals across cases and noise realizations.

## Disclosures

### Funding

The authors declare that no funds, grants, or other support were received for the preparation of this manuscript.

### Competing Interests

Babak Saboury is a cofounder of United Theranostics. Carlos F. Uribe and Arman Rahmim are cofounders of Ascinta Technologies Inc. Zekun Li was previously an employee of United Imaging Healthcare; this employment had ended at the time of manuscript submission. No other potential conflict of interest relevant to this article was reported.

### Author Contributions

All authors contributed to the conceptual development and refinement of the EQUIP-RPT framework through regular group discussions. Z.L. and A.K.J. led the overall drafting of the manuscript. All authors contributed to writing, reviewed and revised the manuscript critically for important intellectual content, approved the final version, and agree to be accountable for the work.

### Data Availability

No datasets were generated or analyzed during the current study.

### Ethics approval

Not applicable.

### Consent to participate

Not applicable.

### Consent to publish

Not applicable.